\documentclass[colorlinks]{goose-article}

\usepackage{threeparttable}

\newcommand{\vs}{\emph{vs} }

\newcommand{\ie}{i.e.\ }

\newcommand{\Lstar}{L^*}
\newcommand{\sigmaf}{\sigma_\mathrm{f}}
\newcommand{\sigmastar}{\sigma^*}
\newcommand{\sigmamin}{\sigma_{\min}}
\newcommand{\sigmac}{\sigma_c}
\newcommand{\expnu}{\nu}

\newcommand{\expdf}{d_f}

\newcommand{\expkappa}{\tau_A}
\newcommand{\ellblock}{\ell_0}
\newcommand{\ellblocktilde}{\tilde{\ell}_0}

\hypersetup{pdfauthor={T.W.J. de Geus}}

\title{Scaling theory for the statistics of slip at frictional interfaces}

\author[1]{T.W.J.~de~Geus}
\author[1]{Matthieu Wyart}

\affil[1]{
    Physics Institute,
    \'{E}cole Polytechnique F\'{e}d\'{e}rale de Lausanne (EPFL) Switzerland}

\DeclareSubrefFormat{mybold}{\textbf{{#2}}}
\begin{document}

\twocolumn[
    \begin{@twocolumnfalse}
        \maketitle
        \begin{abstract}
            \noindent
            Slip at a frictional interface occurs via intermittent events.
            Understanding how these events are nucleated, can propagate,
            or stop spontaneously remains a challenge, central to earthquake science and tribology.
            In the absence of disorder, rate-and-state approaches
            predict a diverging nucleation length at some stress $\sigmastar$,
            beyond which cracks can propagate.
            Here we argue for a flat interface that disorder is a relevant
            perturbation to this description.
            We justify why the distribution of slip contains two parts:
            a powerlaw corresponding to `avalanches', and a
            `narrow' distribution of system-spanning `fracture' events.
            We derive novel scaling relations for avalanches,
            including a relation between the stress drop and
            the spatial extension of a slip event.
            We compute the cut-off length beyond which avalanches cannot be stopped by disorder,
            leading to a system-spanning fracture,
            and successfully test these predictions in a minimal model of frictional interfaces.
        \end{abstract}
    \end{@twocolumnfalse}
]
\sloppy

\section{Introduction}

When a frictional interface is driven quasistatically,
periods of loading are punctuated by sudden macroscopic slip events.
Field observations on earthquakes \cite{Rice1993,Scholz1998} and laboratory studies
support that slip nucleates at weak regions of the interface
and then propagates ballistically as a fracture
\cite{Xia2004,Rubinstein2004,Ben-David2010,Passelegue2013,Heaton1990,Zheng1998,
    Roch2022,Svetlizky2014,Svetlizky2016}.
Understanding under which conditions large slip events are triggered and can propagate is
central to tribology, for example to explain the observed variability of friction coefficients
\cite{Ben-David2011,Popov2010b,Rabinowicz1992}.
It is also key for earthquake science \cite{Brace1966}.
Earthquakes are powerlaw distributed when averaged
over many faults \cite{Gutenberg1954}.
When fault specific data are considered,
observations are debated.
Some studies find a bimodal distribution, consisting of a powerlaw behaviour at
small magnitude on several decades,
an absence of events at intermediate magnitude,
and a few top outliers for which the magnitude is large \cite{Wesnousky1994}.
Other studies suggest instead a continuous powerlaw \cite{Page2011}.
This debate is complicated by the fact that an individual fault consists of many segments,
whose length distribution is itself self-similar \cite{Manighetti2009}.
Here we propose an explanation for the bimodal distribution of slip events when slip occurs
at a single interface, which we consider to be disordered but essentially flat.

These questions are complicated by the fact that frictional forces
can decrease with sliding velocity.
Various mechanisms can lead to such a velocity-weakening, including thermal creep
\cite{Scholz1976,Baumberger2006,Rabinowicz1956a,Marone1998,Heslot1994,Vincent-Dospital2020}
or the mere effect of inertia
\cite{Fisher1997,Ramanathan1997,Schwarz2001,Salerno2012a}.
Rate-and-state models \cite{Dieterich1979,Rice1983,Ruina1983,Scholz1998} describe the dynamics of
frictional interfaces via differential equations that capture velocity weakening.
The latter is characterised by a length scale $L_c$ below which
its effect is small in comparison to elastic forces
\cite{Lebihain2021,Perfettini2003,Ray2017,Dublanchet2018,Albertini2021,Schar2020}.
Importantly, in the case where the stress as a function sliding velocity displays a minimum
$\sigmamin$, this approach predicts \cite{Zheng1998,Ohnaka1990,Brener2018} a characteristic
stress $\sigmastar$ very close to $\sigmamin$, beyond which a slip pulse of spatial extension
larger than $L^*$ will invade the system.
In \cite{Brener2018}, it is found that
$\Lstar/L_c \sim (\sigma - \sigmastar)^{- 1}$.
Yet, these results apply when the interface
is homogeneous: their validity in the presence of disorder
nor their connection to the observed broad distribution of earthquakes is clear.

Another approach describes how an elastic manifold driven through a
disordered medium can be pinned by disorder \cite{Fisher1983,Fisher1998},
and was specifically applied to frictional interfaces
\cite{Fisher1983,Fisher1997,Fisher1998,Ramanathan1997}.
In simple settings that exclude the existence of velocity weakening,
the stress of a quasistatically driven interface converges to
some critical value, where slip events are powerlaw distributed.
Unfortunately, these results do not apply in presence of velocity-weakening effects where even
the presence of large avalanches was debated
\cite{Fisher1997, Dahmen1998,Schwarz2003,Maimon2004}~\footnote{
    In this article we define \emph{avalanches} as a cascade of slip events
    that does not involve the entire system.
}, yet experimentally observed in \cite{Baldassarri2006}.
Very recently \cite{deGeus2019},
we introduced a minimal model of frictional interfaces that contains long-range
elastic interactions, disorder, and inertia.
Criticality was observed, with powerlaw avalanches
whose size can span four decades as the stress reaches some critical value $\sigmac$.
Yet, inertia introduces novel phenomena.
For example, in a finite system the distribution of events
is bimodal: powerlaw distributed avalanches co-exist with system-spanning events.
Which mechanism causes such large avalanches,
and how their duration, length scale, and stress drop are related to each other remain unknown.
So is the relationship between $\sigmac$ governing avalanches and rate-and-state approaches.

In this article, we argue theoretically that $\sigmac = \sigmastar$, implying that rate-and-state
approaches capture the critical stress affecting the slip statistics.
Yet, we find that disorder is a relevant perturbation:
consequently, previous results for the diverging nucleation length scale
near $\sigmac$ neither based on a homogeneous system \cite{Brener2018}
nor on Griffith's argument \cite{deGeus2019} apply.
Our current analysis justifies the presence of large powerlaw avalanches
and leads to scaling relations between their length, stress drop, and duration;
which are found to be related to a fractal property of the slip geometry at the interface.
We successfully test all predictions numerically in the minimal model of \cite{deGeus2019}.

\section{Scaling theory for slip events with velocity weakening}

\paragraph{Observables describing slip events}

\begin{figure}[htp]
    \centering
    \subfloat{\label{fig:theory:PS}}
    \subfloat{\label{fig:theory:flow}}
    \includegraphics[width=\linewidth]{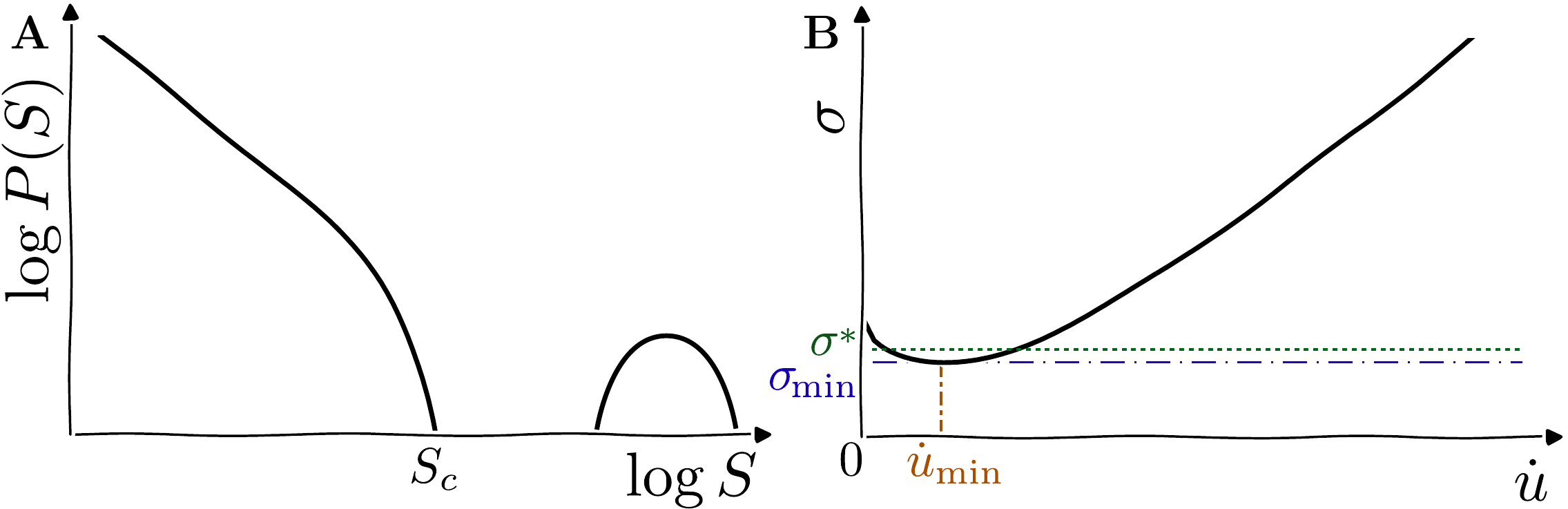}
    \caption{
        \textbf{\protect\subref*{fig:theory:PS}.}
        Sketch of distribution of avalanche sizes in a finite system:
        on the left there is a powerlaw distribution cut-off
        beyond some characteristic value $S_c$,
        on the right there are system-spanning events.
        \textbf{\protect\subref*{fig:theory:flow}.}
        Sketch of a flow curve
        (stress $\sigma$ \vs slip rate $\dot{u}$)
        with a minimum at $\sigma = \sigma_{\min}$.
        (with corresponding slip rate $\dot{u}_{\min}$).
    }
\end{figure}

We characterise slip events by
their linear spatial extension (or `width' in one dimension) $A$,
the total slip (the increment of slip, or increment of displacement discontinuity, integrated
across the event's `width') $S$, and their duration $T$.
As sketched in \cref{fig:theory:PS} and as reported in some systems displaying stick-slip
\cite{Wesnousky1994,Schwarz2001,deGeus2019},
the distribution $P(S)$ consists of two parts.
First, there is a powerlaw distribution cut-off beyond some characteristic value $S_c$,
\ie $P_a(S) = S^{-\tau} f(S / S_c)$ where $f$ is a rapidly decreasing function of its argument.
We call the associated events ($S < S_c$) `avalanches' and denote
by $A_c$ their cut-off spatial extension along the interface.
Second, there are system-spanning slip events of extension $A \approx L$, where $L$ is the
system size, resulting in the `bump' at large $S$ in \cref{fig:theory:PS}.
Empirical observations \cite{Gutenberg1944,Kagan2014}
support the existence of scaling behaviours for avalanches:
\begin{align}
    \label{eq:PS}
    P(S) &\sim S^{-\tau}, \\
    \label{eq:S}
    S &\sim A^{\expdf}, \\
    \label{eq:T}
    T &\sim A^z.
\end{align}
A fourth scaling relation was instead observed
in the simple model of interface of \cite{deGeus2019}:
\begin{equation}
    \label{eq:Ac}
    A_c \sim (\sigma - \sigma_c)^{- \expnu}.
\end{equation}

\paragraph{Observables describing the static interface}

We consider an interface that is overall flat
and homogeneously loaded.
Disorder can be exogenous, stemming for example from asperities on the surfaces of the two bodies,
or instead be endogenous and result from the history from previous slip events
that lead to irregular stresses along the interface.
Upon loading, the interface will acquire some slip $u(r)$ at location $r$.
Due to the disorder, $u(r)$ will fluctuate spatially.
These fluctuations can be characterised by
introducing the roughness exponent $\zeta$ of the interface \cite{Fisher1998}:
\begin{equation}
    \label{eq:zeta}
    ||u(r) - u(r')|| \sim ||r - r'||^\zeta
\end{equation}
with $|| \ldots ||$ the root-mean-square.

We have so far introduced five exponents: $\tau$, $\zeta$, $\expnu$, $z$, $\expdf$.
Our central goal is to propose three new scaling relations
relating $\zeta$, $\expnu$, $z$, $\expdf$ together, allowing for a stringent empirical test of our
views.

\paragraph{Effect of disorder on the rate-and-state description}

Previous attempts to describe the joint effects of disorder and
velocity-weakening sought to treat the latter as a perturbation \cite{Fisher1997,Dahmen1998}.
We take the opposite approach, and seek to characterise how disorder
affects the dynamics of a homogeneous interface subjected to velocity weakening,
as captured by the rate-and-state description \cite{Zheng1998,Brener2018}.
The relationship $\sigma(\dot{u})$ between the far-field stress $\sigma$
and the slip rate $\dot{u}$, at any location, is key in this approach.
If it does display a minimum $\sigmamin$ for some slip rate $\dot{u}_{\min}$
as illustrated in \cref{fig:theory:flow},
then it was shown that beyond some stress $\sigmastar$ just above $\sigmamin$,
slip events of length $\Lstar \sim (\sigma - \sigmastar)^{-1}$
can nucleate system-spanning events \cite{Brener2018}.

However, where it makes sense for a homogeneous system to consider $\sigmastar$
as a quantity that does not vary in space,
in a disordered system its structure is locally random.
A patch of material of linear extension $A$
can still be described by some effective threshold $\sigmastar(A)$,
but this quantity must vary in space.
$\sigmastar(A)$ thus display fluctuations, whose magnitude we denote $\delta \sigmastar (A)$.
They can only disappear in the thermodynamic limit $A\rightarrow \infty$ where randomness
self-averages and homogenisation is achieved.
In general one expects:
\begin{equation}
    \label{eq:delta_sigma}
    \delta \sigmastar \sim A^{-\chi}.
\end{equation}
Classical arguments based on disorder imply $\chi \leq (d+\zeta) / 2$ \cite{Chayes1986,Fisher1998}
\footnote{ When a portion of linear length $A$ of the interface moves, it will
explore a new realisation of the disorder.
If the disorder is assumed to have no spatial correlations, that motion will be affected by
$N_r={\cal O}(S)$ random numbers, where $S$ is the integrated slip.
We shall see below that $S$ follows $S\sim A^{d+\zeta}$.
From the central limit theorem, any threshold characterising motion cannot be defined with a
precision finer than $1/\sqrt{N_r}$, leading to the bound stated in the main text.
}.
Here, $d$ is the dimension of the interface
(separating objects of dimensions $d + 1$).
Below, we will provide data supporting that this bound is not saturated.

If $\chi\leq 1$ (as we shall confirm empirically for $d = 1$),
we now argue that due to these fluctuations,
rate-and-state results on nucleation in homogeneous systems cannot apply to disordered ones.
Indeed, consider $\sigma - \sigmastar$ to be small but positive,
and a slip event occurring on a length scale $\Lstar \sim (\sigma - \sigmastar)^{- 1}$.
On that length scale,
the fluctuations of $\sigmastar$ are stronger than the distance
to threshold $\sigma - \sigmastar$ when the latter is small:
$\delta \sigmastar (\Lstar) \sim (\Lstar)^{-\chi}
    \sim (\sigma - \sigmastar)^{\chi} \geq (\sigma - \sigmastar)$.
Thus this theory neglecting the fluctuations of $\sigmastar$
cannot self-consistently hold near threshold.

\paragraph{Roughness of the interface}

As discussed above, the strength $\sigmastar(A)$ of a patch of size $A$ varies in space.
The interface must adjust to these variations:
the slip $u(r)$ will be larger at locations $r$ where $\sigmastar(A)$ is small.
Fluctuations of elastic stresses follow fluctuations of strain, which between two points $r$ and
$r'$ are of order ${||u(r) - u(r')|| / ||r - r'||}$.
If $||r - r'||\sim A$, using \cref{eq:zeta} these fluctuations are of order $A^{\zeta-1}$.
As is more generally the case for an elastic manifold in disordered
environments \cite{Fisher1998}, we expect such adjustments of the interface slip to stop when
these fluctuations of elastic stresses
are of order of the fluctuations of $\sigmastar(A)$ on that scale
(of order $A^{-\chi}$, see \cref{eq:delta_sigma}), leading to:
\begin{equation}
    \label{eq:chi}
    \chi = 1 - \zeta.
\end{equation}

\paragraph{Justifying powerlaw avalanches}

We now argue that \cref{eq:delta_sigma} gives a natural explanation for the presence
of powerlaw slip events or `avalanches'.
Consider a system at $\sigma = \sigmastar$ where a slip starts to occur at the
origin, whose extension grows in time as $A(t)$.
During this process, the system encounters new realisations of the disorder,
and also explores larger regions of space.
Thus, the effective threshold $\sigmastar(A(t))$ for slip propagation felt
on that scale will vary in time.
The dynamics will stop if it becomes larger than the applied stress,
\ie ${\sigmastar(A(t)) > \sigmastar = \sigma}$.

Following the depinning literature, simple arguments then constrain the statistics of stopping
events \cite{Fisher1998}.
$\sigmastar(A(t))$  can be thought as a random variable that evolves continuously around its mean $\sigmastar$. $\sigmastar(A(t))$ will lose memory of its current value when the patch size $A(t)$ increases significantly, i.e.\ $\sigmastar(A(t_1))$ at time $t_1>t_0$ decorrelates from $\sigmastar(A(t_0))$ when $A(t_1)-A(t_0)\geq A(t_0)$.
As a result, every time $A$ doubles in size, there is a finite
probability $p_2$ that $\sigmastar(A(t))-\sigmastar$ has changed sign, and that slip has stopped.
Such a property implies a powerlaw distribution $P(A) \sim A^{- \expkappa}$ with
$\expkappa = 1 - \ln(1 - p_2) / \ln(2)$ \footnote{
    Such a property reads $[\phi(A) - \phi(2A)] / \phi(A) = p_2$, where $\phi(A)$ is
    the cumulative distribution characterising the probability that slip is larger than $A$,
    $\phi(A) = \int_{y > A} d y \big( P(y) \big) \sim A^{1 - \tau_A}$.
}.
Thus we predict that the stress $\sigma_c$ at which avalanches are
powerlaw, follows:
\begin{equation}
    \label{eq:sigmac}
    \sigma_c = \sigmastar.
\end{equation}

\paragraph{Maximal avalanche extension $A_c$}

\begin{figure}[htp]
    \centering
    \includegraphics[width=\linewidth]{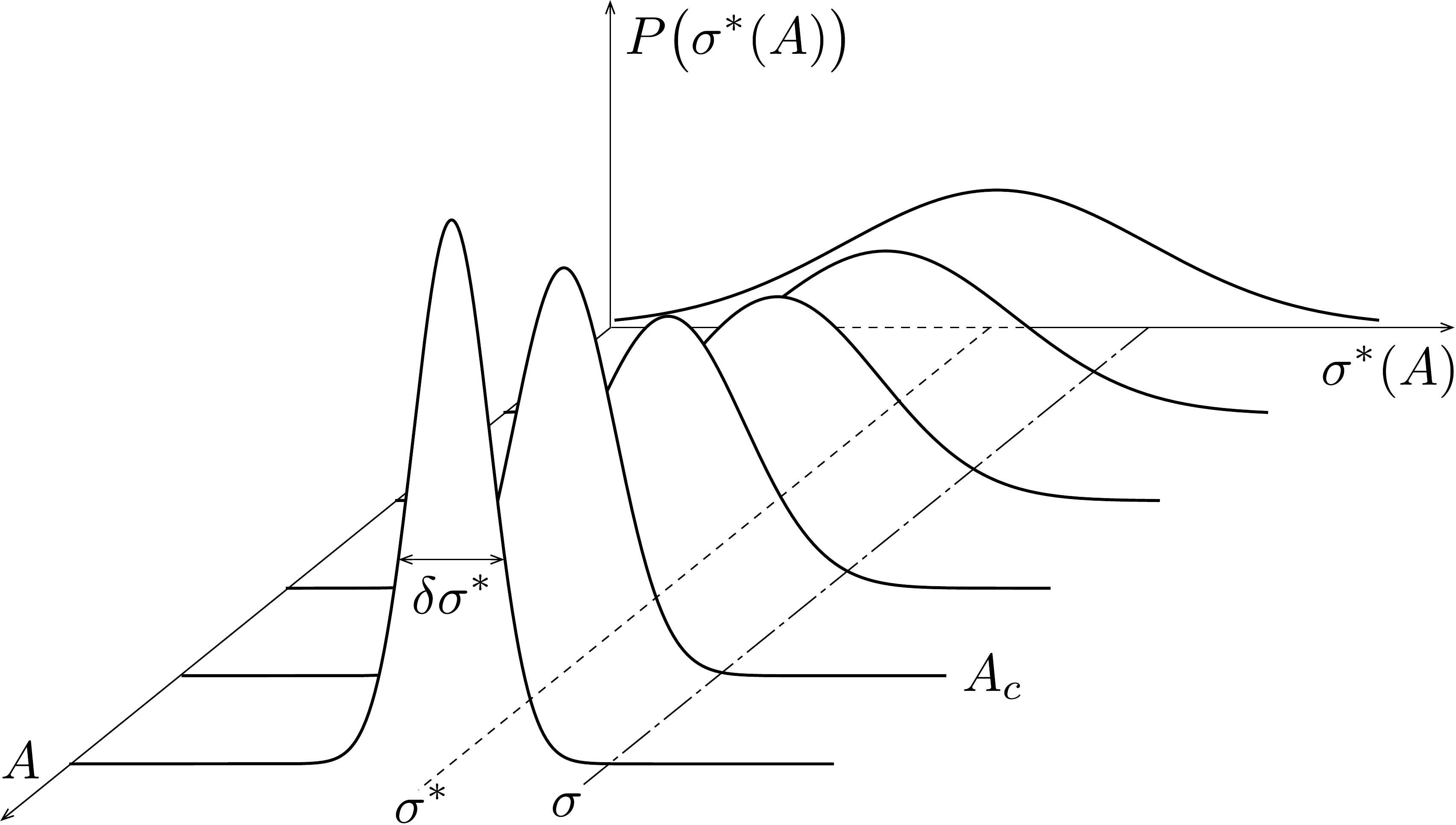}
    \caption{
        (Sketch) A patch of size $A$ has a strength $\sigmastar(A)$,
        which, due to disorder, is distributed around its thermodynamic value $\sigmastar$.
        Denoting $\delta \sigmastar (A)$ the width of this distribution for a given size $A$,
        $\delta \sigmastar$ will decrease as $A$ increases --
        a large patch has a better estimate of the true, thermodynamic, $\sigmastar$.
        Consequently, if an event is nucleated at $\sigma > \sigmastar$,
        disorder can still stop the event if its width $A$ is sufficiently small
        ($\delta \sigmastar$ is large for small $A$, so there is a finite probability that
        for that region in space $\sigmastar(A) > \sigma$).
        However, an event of size $A > A_c$ cannot be stopped by disorder, as the probability
        that that region in space has a strength $\sigmastar(A) > \sigma$ vanishes.
    }
    \label{fig:distro_sigma}
\end{figure}

Consider the same argument applied to the case $\sigma > \sigmastar$.
As long as the scale of fluctuations $\delta \sigmastar(A) \gg \sigma - \sigmastar$,
the difference $\sigma - \sigmastar$ is insignificant (as sketched in \cref{fig:distro_sigma}),
and one recovers a powerlaw distribution of slip events as argued above.
However, in the other limit where $\delta \sigmastar(A) \ll \sigma - \sigmastar$,
one always has $\sigmastar(A) < \sigma$.
In that regime, disorder cannot stop a propagating `crack':
disorder is irrelevant, and the interface can be safely approximated to be homogeneous.
Rupture is then predicted to be correctly described by homogeneous rate-and-state laws.

The crossover between these two regimes occurs for a slip
extension $A_c$ satisfying $\delta \sigmastar(A_c) \sim \sigma - \sigmastar$.
Using \cref{eq:delta_sigma,eq:zeta} one obtains
$A_c \sim (\sigma - \sigma_c)^{- \expnu}$ (\cref{eq:Ac}) with:
\begin{equation}
    \label{eq:nu}
    \nu = \frac{1}{\chi} = \frac{1}{1 - \zeta}
\end{equation}
which corrects a Griffith argument proposing $\nu = 2$ \cite{deGeus2019},
which neglected the (dominant) effect of disorder.

\paragraph{Geometry of avalanches}

When slip occurs on a length scale $A$, the disorder characterising
this region evolves.
Locally, the interface strength can decrease by some increment
$\delta \sigmastar(A) \sim A^{-\chi}$.
Slip will stop when the local stress, proportional to $u / A$, decreases by a similar
magnitude.
It corresponds to a slip of order $u$ satisfying $u / A \sim A^{-\chi}$.
Using that by definition $S \sim A^d u$ then leads to $S \sim A^{\expdf}$ (\cref{eq:S}) with:
\begin{equation}
    \label{eq:df}
    \expdf = d + 1 - \chi = d + \zeta.
\end{equation}

Note that \cref{eq:chi,eq:nu,eq:df} are well known to hold
in the absence of inertia and velocity weakening \cite{Fisher1998}.
The proposition that they describe the pinning of velocity-weakening
elastic materials, where avalanches co-exist with system-spanning events and where
the flow curve has a minimum at finite slip rate, is to the best of our knowledge new.
Indeed, most previous theoretical works
argued that powerlaw avalanches would be absent in that case \cite{Fisher1997,Dahmen1998}.
Yet, the values of the exponents will differ in the absence or presence of large inertia,
as we document below.
We now turn to a scaling relation that is specific to the presence of velocity weakening.

\paragraph{Duration of avalanches}

For stresses in the vicinity of $\sigmastar$, according to the flow
curve sketched in \cref{fig:theory:flow},
slip is possible only if the slip rate lies in the vicinity of $\dot{u}_{\min}$.
We make the hypothesis that within an avalanche,
a sizeable fraction of the interface is slipping at any given point in time.
The characteristic slip rate of an avalanche, $\dot{u}$, thus satisfies
$\dot{u} \equiv S / (A^d T) \sim \dot{u}_{\min}$
that behaves as a constant as $\sigma \rightarrow \sigmastar$,
which implies $T \sim A^z$ (\cref{eq:T}) with:
\begin{equation}
    \label{eq:z}
    z = \expdf - d = \zeta
\end{equation}

\section{Testing the theory}

\subsection{A Rosetta Stone Model for frictional interfaces}

We consider the minimal model of frictional interface containing disorder,
long-range elasticity and inertia introduced in \cite{deGeus2019}.
Its details, as well as the dimensionless units we choose,
are reviewed in the appendix.
As illustrated in \cref{fig:model}, the frictional interface is
discretised in $L$ (orange) ``blocks'' of unit size.
Such a mesoscopic description is standard in Burridge-Knopoff type models \cite{Burridge1967}
or in the depinning literature \cite{Fisher1997}.
In the absence of inertia, it can successfully describe interfaces beyond the so-called
``Larkin length'' \cite{Cao2018},
below which asperities always collectively rearrange and the details of the disorder matters.
In the presence of velocity weakening, such a description can describe the interface beyond
another length scale $L_c$,
below which elastic forces dominate those stemming from velocity weakening
\cite{Lebihain2021,Perfettini2003,Ray2017,Dublanchet2018,Albertini2021,Schar2020,Uenishi2003}.
Below, we estimate $L_c$ to be about 20 blocks in our model.

The interface is embedded within two homogeneous linear elastic bodies,
of total height $\approx L$,
modelled by finite elements (blue) \footnote{
    Note that in our model, friction-like properties emerge from the presence of
    disorder and inertia.
    These properties are not prescribed form the start as in block-spring models \cite{Langer1996}.
}.
The system is driven at the top and fixed at the bottom,
and presents periodic boundaries on the horizontal axis.
Each block responds linear elastically up to a randomly chosen yield stress, whereupon it slips.
This corresponds to a potential energy that, as a function of local slip,
comprises a sequence of parabolic wells of random width, as illustrated in \cref{fig:model}.
Disorder stems from randomly choosing the yield stresses,
which are proportional to the width of the wells.
In the absence of inertia,
such models are used to study the depinning transition \cite{Jagla2017}, where they allow for fast
simulations and a simple definition of avalanches, whose size $S$ is simply the number of times
blocks rearranged within an event.

\begin{figure}[htp]
    \centering
    \subfloat{\label{fig:model}}
    \subfloat{\label{fig:typical}}
    \includegraphics[width=\linewidth]{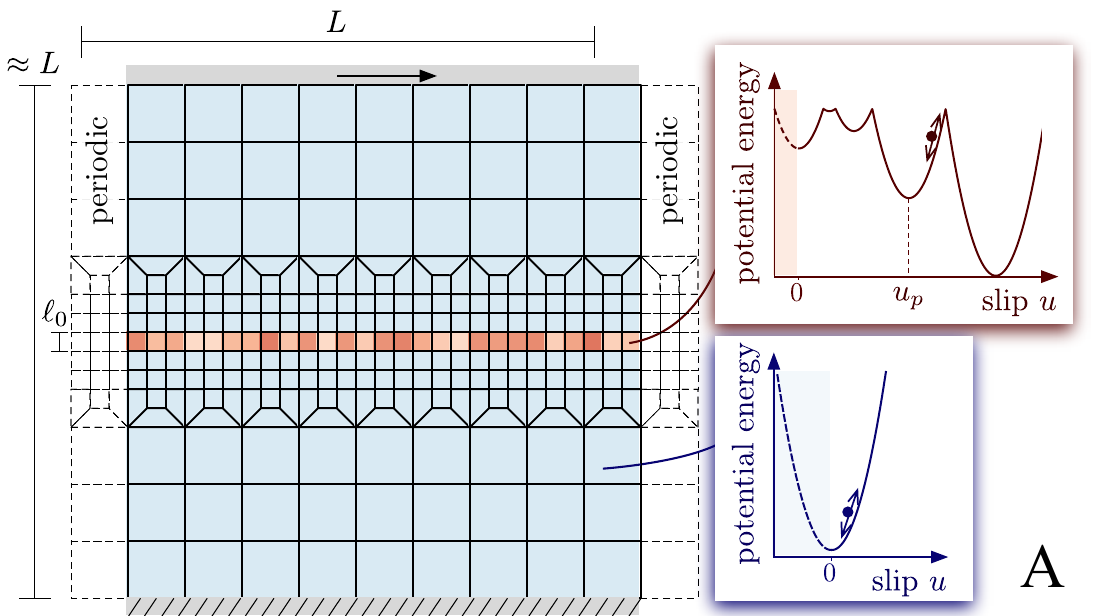}
    \\
    \includegraphics[width=.8\linewidth]{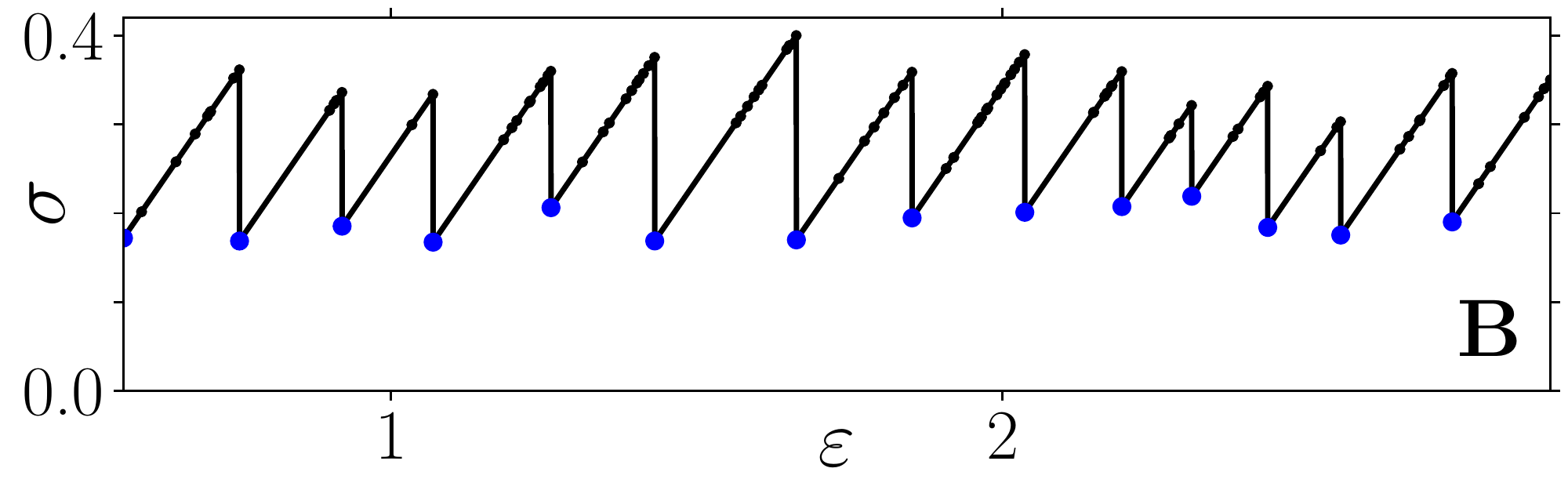}
    \caption{
        \textbf{\protect\subref*{fig:model}.}
        The frictional interface is modelled
        using $L$ `blocks' (finite elements in orange).
        The interface is embedded between two elastic bodies
        (also discretised using finite elements in blue)
        such that the entire system is approximately square.
        The potential energy of blocks along the interface and in the bulk
        are sketched as indicated.
        \textbf{\protect\subref*{fig:typical}.}
        Macroscopic stress $\sigma$ \emph{vs} strain $\varepsilon$
        response to a quasistatic drive in which the top boundary is displaced `infinitely' slowly.
        After each vanishingly small step, energy is minimised by following the inertial dynamics.
        The position of the top boundary and the reaction force yield $\varepsilon$ and $\sigma$
        (see details about units in the appendix).
        `Events', during which at least on block yields,
        are indicated with a marker, in blue when they are system spanning.
    }
\end{figure}

We consider standard inertial dynamics,
with a small damping term chosen to ensure that elastic waves become damped
after propagating on a length scale of order $\sim L$,
modelling the leakage of heat at the system boundary.
As we show below, the presence of (weakly damped) inertia leads to a velocity weakening,
well fitted by rate-and-state description.
Thus, this model is ideally suited to build
a dictionary between the rate-and-state description (that focuses on velocity-weakening) and
the depinning viewpoint (that focuses on disorder).

\begin{figure}[htp]
    \centering
    \subfloat{\label{fig:eventmap:a}}
    \subfloat{\label{fig:eventmap:b}}
    \includegraphics[width=\linewidth]{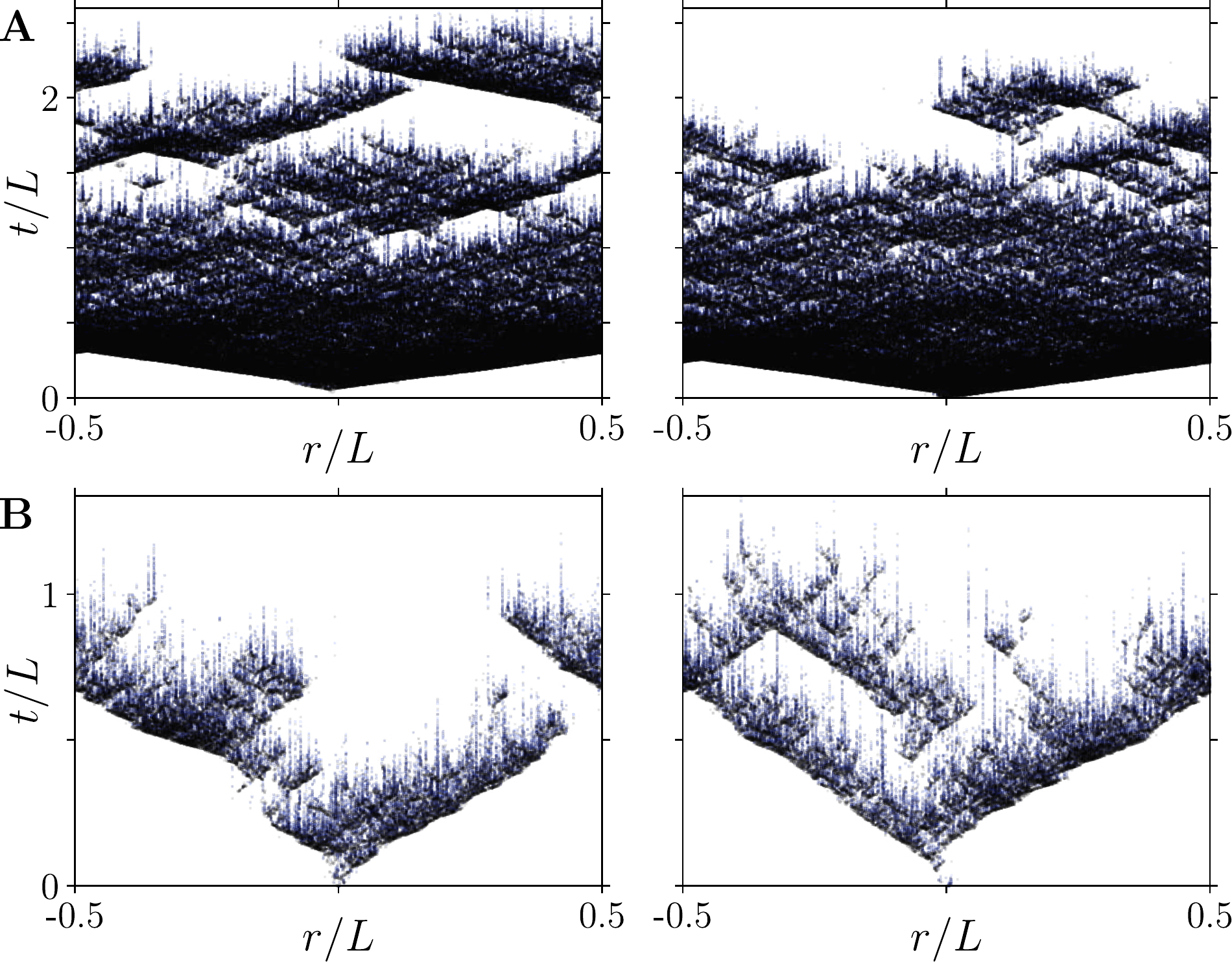}
    \caption{
        Event maps for two representative system-spanning events (\protect\subref*{fig:eventmap:a})
        and two large avalanches nucleated at $\sigma_c$ (\protect\subref*{fig:eventmap:b}).
        A point is placed in time $t$ and space $r$ for each yield event
        (black if the block yields in the positive direction and
        blue if it yields in the negative direction).
    }
    \label{fig:eventmap}
\end{figure}

\subsection{Calibrating and testing rate-and-state}

\paragraph{Stationary velocity weakening}

\begin{figure}[htp]
    \centering
    \includegraphics[width=\linewidth]{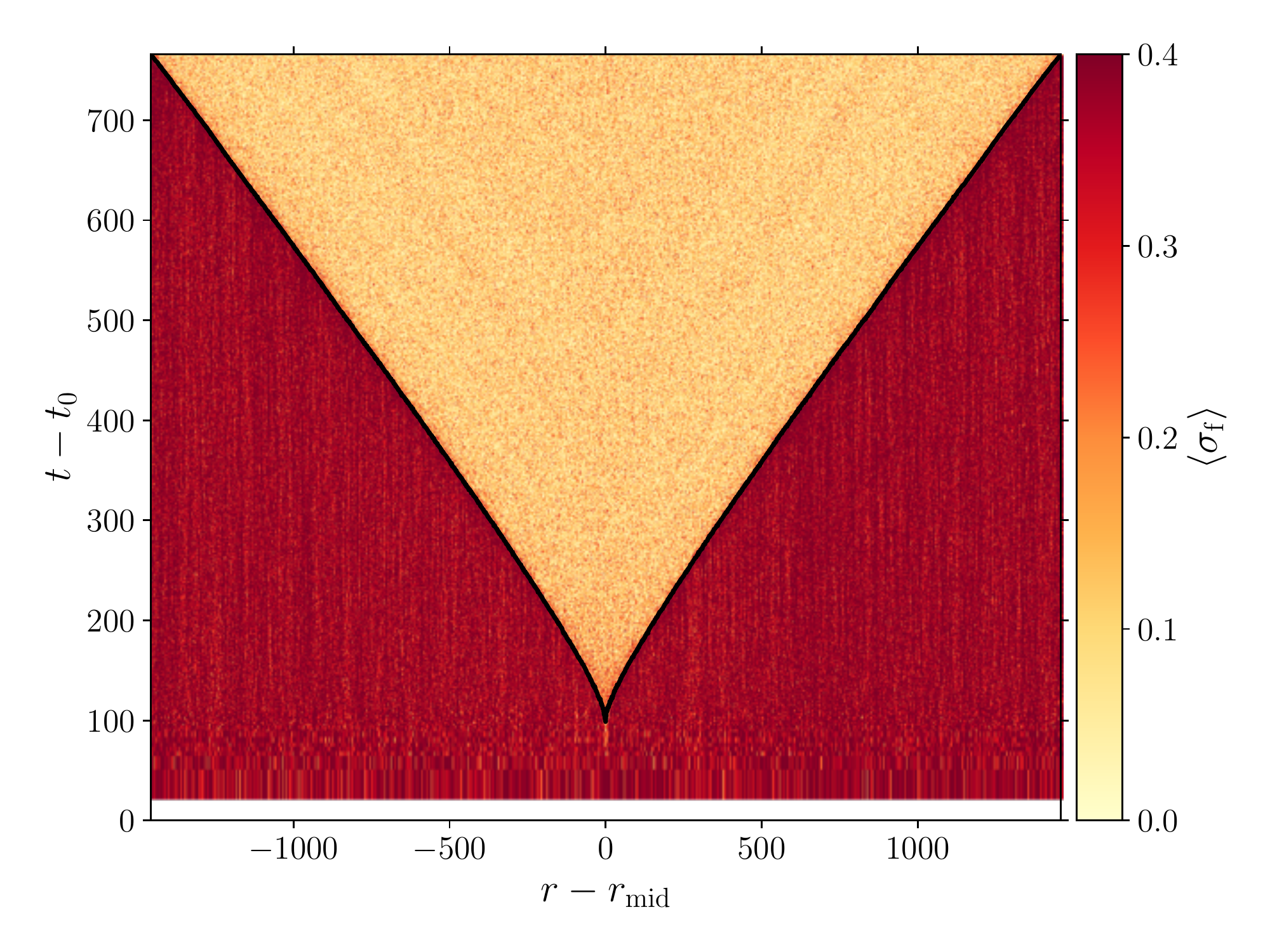}
    \caption{
        Ensemble average stress $\sigmaf (r, t)$ (colour) along the weak layer $r$ (horizontal axis)
        as a function of time since the beginning of the event (vertical axis).
        The average is taken with respect to the (in general time-dependent)
        center of the event (details in appendix).
        Identically, the average number of times a block yielded, $s(r, t)$, is recorded,
        whose $s(r, t) = 1$ contour is shown to delimit the front of the event.
    }
    \label{fig:map}
\end{figure}

Velocity-weakening is already apparent under a quasistatically imposed shear,
where it leads to stick slip.
As illustrated in \cref{fig:typical},
system-spanning events drop the stress to some value indicated in blue,
are punctuated by `avalanches' in which a fraction of the blocks yield (small markers).
\cref{fig:eventmap:a} shows a spatiotemporal map of two system-spanning events
spontaneously occurring upon loading (\ie at large stress),
whereas \cref{fig:eventmap:b} illustrates two large avalanches
(that we triggered directly after system spanning events, \ie at low stress).
For events nucleated at high stress, we show the average stress along the interface as a function
of time in \cref{fig:map} \footnote{
    For completeness: we average on $40$ system spanning events, triggered in the highest bin of
    \cref{fig:PhiA,fig:nu,fig:PhiA_collapse}.
    These results are representative of system spanning that nucleate spontaneously.
}.
We use a blue contour to delimit the (average) event,
using which we see that the stress drops significantly inside the event.
Note that, in this work, we do not focus on the properties of the front
(that travels at a velocity higher than the shear wave speed).

To describe the observed velocity-weakening,
we consider the rate-and-state description \footnote{
    We emphasise that we include only elasticity, local yielding,
    and inertia in our numerical model.
    We \emph{do not} impose the rate-and-state model.
    Rather it describes the emerging properties of the interface well.
} relating the
interfacial stress $\sigmaf$ as a function of slip rate $\dot{u}$ and time~$t$ as:
${\sigmaf = \sigma_s + a \ln ( \dot{u} ) + b \ln ( \theta (\dot{u}, t) )}$.
Here, the time dependence enters implicitly through a `state' parameter $\theta$.
Furthermore, $\sigma_s$ is some offset and $a$ and $b$ are parameters.
Usually, the state parameter is assumed to follow a
simple linear ageing law $\dot{\theta} = 1 - \theta \dot{u} / D_c$.
This equation captures that memory is lost once slip becomes larger than a distance $D_c$,
beyond which the steady state ($\dot{\theta} = 0$) is reached,
which implies the stationary behaviour:
\begin{equation}
    \label{eq:rate-and-state}
    \sigmaf = \sigma_s + (a - b) \ln ( \dot{u} ).
\end{equation}
Note that in frictional experiments,
the ``state'' parameter is often associated to the real contact area.
This is not the case in our model, where the contact area is fixed.
Instead,
we think of the state parameter as characterising the mechanical noise stemming from inertia,
that must take a finite time to reach a stationary equilibrium.

\begin{figure}[htp]
    \centering
    \includegraphics[width=\linewidth]{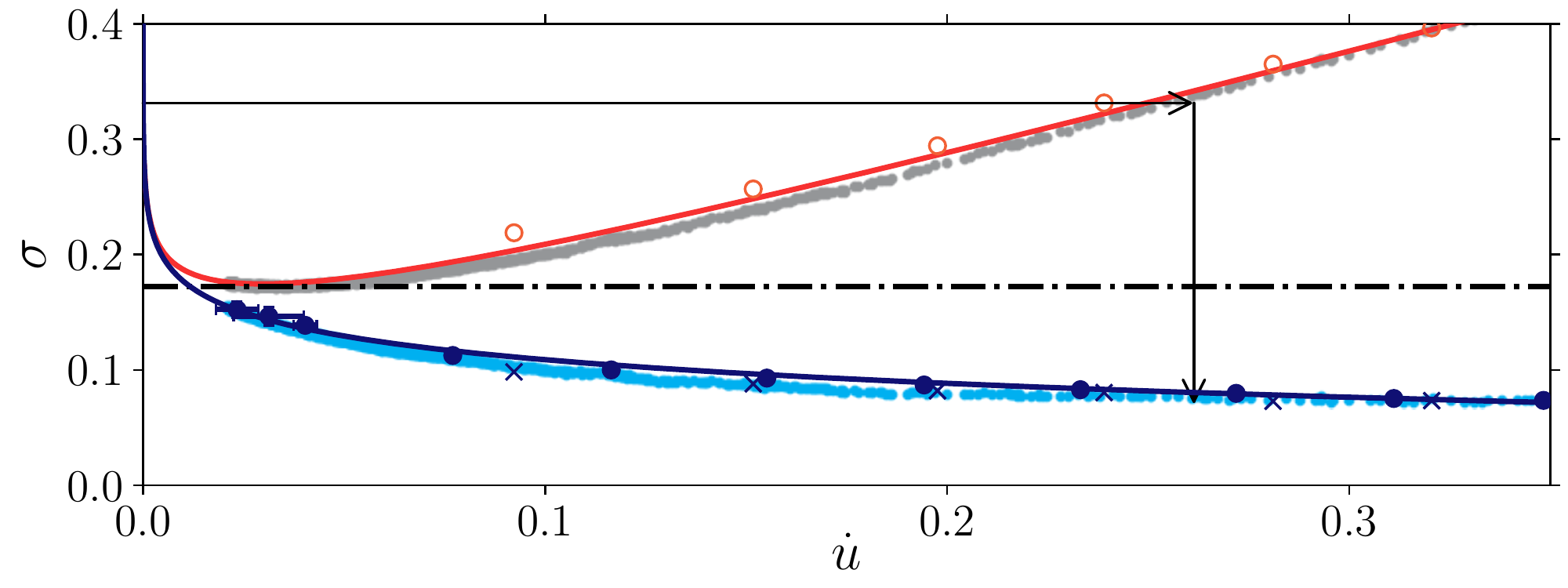}
    \caption{
        In blue (solid markers): Constitutive behaviour of the
        interfacial stress $\sigmaf$ \vs slip rate $\dot{u}$.
        The ``rate-and-state'' law, \cref{eq:rate-and-state}, is fitted on the solid blue markers
        that are measured in the steady state of a ``flow experiment''
        in which the system is subjected to a constant shear rate.
        The fit is confirmed through a measurement of $(\sigmaf, \dot{u})$
        during system-spanning events, as they stop ($\dot{u}$ decreases toward zero),
        as shown using light-blue points,
        truncated at low $\dot{u}$ when there are no more plastic events.
        In red (open markers): effective flow curve relating the far-field stress $\sigma$
        to the local slip rate $\sigma = \sigmaf(\dot{u}) + \dot{u}$.
        The solid red curve corresponds to the fit obtained from
        the solid blue curve, while grey points are obtained from light-blue points.
        The cross-markers and open markers indicate respectively the measurements of
        $(\sigmaf, \dot{u})$ and $(\sigma, \dot{u})$
        obtained by measuring the slip rate $\dot{u}$ and
        interfacial stress $\sigmaf$ within a rupture occurring at an imposed stress $\sigma$.
        The minimum of the effective flow curve $\sigma_{\min}$ is indicated using a dash-dot line.
    }
    \label{fig:scenario}
\end{figure}

To calibrate \cref{eq:rate-and-state}, we measure the steady state interfacial stress $\sigmaf$
for different imposed slip rates $\dot{u}$, averaged in both space and time.
Measurements corresponds to the solid blue markers in \cref{fig:scenario}.
The solid blue line fits them
according to \cref{eq:rate-and-state}, and leads to $a - b \simeq -0.03$
(and $\sigma_s \simeq 0.041$).

The dynamics are intermittent at imposed $\dot{u} < \dot{u}_{\min}$
(defined in \cref{fig:theory:flow}) \footnote{
    Above, but close to $\dot{u}_{\min}$ intermittency still obstructs a truly
    steady-state measurement.
    We thus measure the instantaneous stress and slip rate along the interface
    leading to the error bars at low $\dot{u}$.
}, and stick slip occurs.
The response $\sigmaf(\dot{u})$
at small strain rate can be estimated by measuring evolution of the spatial average
stress $\sigmaf$ and the slip rate $\dot{u}$ along the weak layer in system-spanning events.
After these events span the system, the average slip rate $\dot{u}$ slowly relaxes toward zero.
These measurements corresponds to the light-blue points in \cref{fig:scenario}
that also are well fitted by \cref{eq:rate-and-state} with the same parameters.

As mentioned in the introduction,
in rate-and-state descriptions there is a characteristic length scale $L_c$
beyond which velocity-weakening effects are important.
As recalled in the appendix, we can now estimate $L_c$ using the value of $a - b$ and a natural
estimate for the slip $D_c$ where stationarity is reached.
We assume that $D_c$ corresponds to the characteristic slip
length for which plasticity occurs in a given block, i.e.\ the typical slip for which one exits the
parabolic potential in \cref{fig:model}.
With this, we obtain $L_c \approx 20$ blocks.
In what follows, we focus on the quantification of slip events that are larger than $L_c$.

\paragraph{Radiation damping leads to a non-monotonic effective flow curve}

The blue curve in \cref{fig:scenario} describes a stationary situation.
However, when a slip event occurs
and has not yet spanned the entire system, it must accelerate
the elastic material around it.
This phenomenon must obviously occur here as well, since we realistically describe the
elastic material around the interface.
\citet{Zheng1998} show that it is captured by a ``radiation damping'' term,
describing the difference between the stress in the far field $\sigma$
and that of the interface $\sigmaf$ during an event that is growing in space.
In our dimensionless units (see appendix), it simply reads:

\begin{equation}
    \label{eq:radiation}
    \sigma = \sigmaf(\dot{u}) + \dot{u}.
\end{equation}

We show the far-field stress $\sigma$ in red in \cref{fig:scenario}.
A key observation is that this curve is non-monotonic,
and presents a minimum $\sigma_{\min} \approx 0.17$.
Following previous works \cite{Zheng1998,Brener2018},
nucleation of a system-spanning, crack-like, event in homogeneous systems is possible
beyond some stress $\sigma^* \approx \sigma_{\min}$.
Below, we will support empirically our prediction that $\sigma^*$ is also
the stress $\sigma_c$ where avalanches display a diverging cut-off.

Another prediction of rate-and-state (less relevant to our present purpose, but useful to
further support the predictive power of rate-and-state in our model)
is that when a system-spanning event starts to invade the material,
away from the rupture front the local stress and slip rate can be
readily extracted from \cref{eq:radiation}, see \citet{Barras2019}.
This result is illustrated using the arrows in \cref{fig:scenario},
showing that for a given imposed stress $\sigma$ (horizontal arrow), $\dot{u}$ and $\sigmaf$ within
system-spanning events can be read from this curve (vertical arrow).
We confirm this construction in \cref{fig:scenario}:
the open markers indicate the applied stress $\sigma$ and the observed slip rate $\dot{u}$.
The cross-markers instead indicate the interfacial stress $\sigmaf$ inside the event
and $\dot{u}$.
We indeed find that $(\sigmaf, \dot{u})$ away from the
rupture front closely follow the identified steady-state flow curve.

\subsection{Statistics of slip events}

We now test the (i) scaling relations we derived earlier for slip
events and (ii) the correspondence between the stress $\sigma_c$ where avalanches diverge and the
rate-and-state characteristic stress $\sigma_{\min}$, in the neighbourhood of which nucleation of
unstable rupture front is predicted in homogeneous systems,
at $\sigmastar \gtrsim \sigmamin$ \cite{Brener2018}.

\paragraph{Interface roughness}

\begin{figure}[htp]
    \centering
    \subfloat{\label{fig:zeta_stress}}
    \subfloat{\label{fig:zeta}}
    \includegraphics[width=\linewidth]{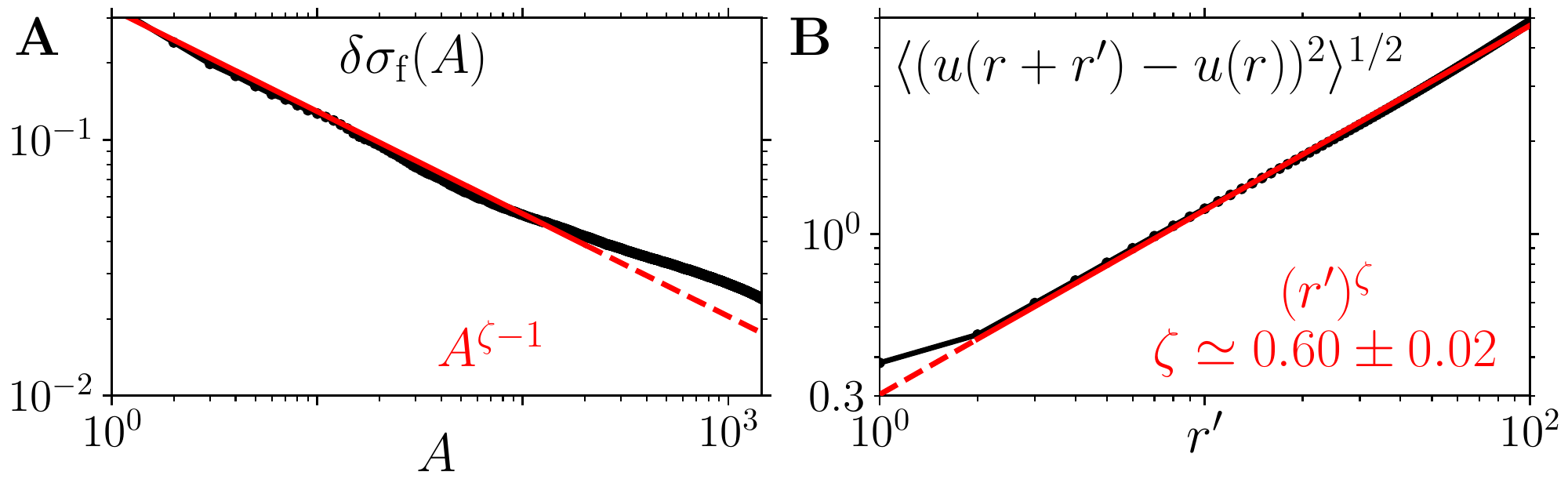}
    \caption{
        \textbf{\protect\subref*{fig:zeta_stress}.}
        Standard deviation of the distribution of mean interfacial stress $\sigmaf$
        on a patch of blocks of size $A$ as a function of $A$.
        In particular, we measure $\text{std}( \{ \sum_{i=j}^{j + A} \sigma_i / A \})$
        (with $i$ the index of a block along the weak layer),
        with the set $\{ \ldots \}$ coming from choosing $j$ at random locations
        in each realisation of the ensemble;
        see appendix for a precise statement.
        \textbf{\protect\subref*{fig:zeta}.}
        Roughness exponent $\zeta$ measured from the mean square fluctuations of slip
        (note that a spatial average on $r$ is implied).
        The error bar on $\zeta$ includes both statistical errors
        (estimated from a least square regression)
        and systematic biases (estimated by reducing the fitting range), as detailed in the
        appendix.
    }
    \label{fig:zeta_all}
\end{figure}

As discussed in the theoretical section, the fluctuation of the strength of the interface
$\delta\sigmastar$ must be reflected
in the fluctuations $\delta\sigmaf$ of the physical stress carried by the interface,
since slip will
tend to occur until the mean stress in a region of size $A$, $\sigmaf(A)$, has relaxed toward
$\sigmastar(A)$.
We have no direct measurement of $\sigmastar(A)$.
Instead, we study $\sigmaf(A)$ and its fluctuation $\delta \sigmaf(A)$.
we consider the system directly after 4000 system-spanning events \footnote{
    Which is about 10\% of the total number of events that occur during quasistatic loading.
}, indicated in blue in \cref{fig:typical}.
We randomly choose patches of linear length $A$ along the interface, and compute the average stress
in each patch.
We then measure the mean and the standard deviation $\delta \sigmaf(A)$ for different length $A$.
In \cref{fig:zeta_stress}, we confirm the powerlaw behaviour
$\delta \sigmaf (A)\sim A^{-\chi}$,
from which we gain an estimate of the exponent $\chi$ entering \cref{eq:delta_sigma}
(assuming $\delta \sigmaf (A) \approx \delta \sigmastar(A)$).

Such stress fluctuations must affect the roughness of the interface, as argued above.
We confirm this view in \cref{fig:zeta}, which displays the relationship between the slip
fluctuations $||u(r) - u(r')||$ and the distance $||r - r'||$.
This observation confirms a powerlaw postulated in \cref{eq:zeta} with an exponent
$\zeta \simeq 0.60$, a measurement consistent
with the scaling relation $\chi={1 - \zeta}$ of \cref{eq:chi}.

\paragraph{Statistics of avalanches}

\begin{figure}[htp]
    \centering
    \subfloat{\label{fig:roughness}}
    \subfloat{\label{fig:PS}}
    \subfloat{\label{fig:SA}}
    \subfloat{\label{fig:TA}}
    \subfloat{\label{fig:PhiA}}
    \subfloat{\label{fig:PhiA_collapse}}
    \subfloat{\label{fig:nu}}
    \includegraphics[width=\linewidth]{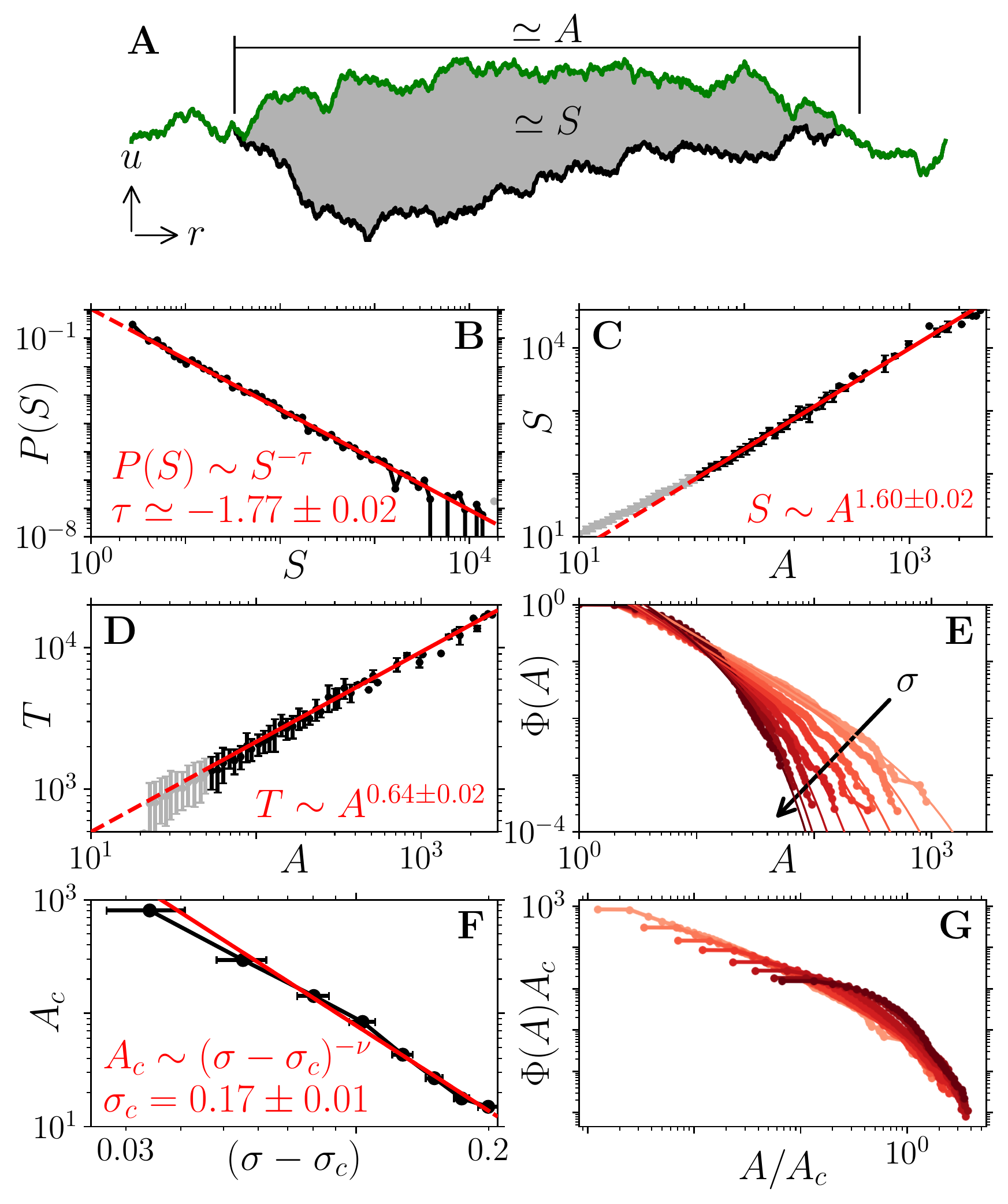}
    \caption{
        \textbf{(\protect\subref*{fig:roughness}-\protect\subref*{fig:TA})}
        \protect\subref*{fig:roughness}.
        Example of slip profile $u(r)$ before (bottom black curve) and
        after (top green curve) an avalanche.
        The linear extension $A$ and total slip $S$ are schematically indicated.
        Our results are consistent with
        $P(S) \sim S^{-\tau}$ (\protect\subref*{fig:PS}),
        $S \sim A^{1 + \zeta}$ (\protect\subref*{fig:SA}), and avalanche duration
        $T \sim A^{\zeta}$ (\protect\subref*{fig:TA}).
        \textbf{\protect\subref*{fig:PhiA}.}
        Cumulative probability $\Phi(A)$ of avalanches (whose $A<L$) triggered at
        different stresses $\sigma$.
        The thin line indicates a fit $\Phi(A) \sim A^{1 - \expkappa} \exp(- A / A_c)$.
        Indeed, the obtained $A_c(\sigma)$ collapses our data (\protect\subref*{fig:nu}).
        \textbf{\protect\subref*{fig:PhiA_collapse}.}
        Fit of the offset $\sigma_c$ such that $A_c \sim (\sigma - \sigma_c)^{- \nu}$,
        while imposing our prediction for $\nu$ in \cref{eq:nu}.
    }
    \label{fig:avalanche}
\end{figure}

To acquire a large statistics, we follow the strategy of \cite{deGeus2019} of manually triggering
$9000$ events such as \cref{fig:eventmap:b}, using a local perturbation at imposed strain,
following a
system spanning event.
The post mortem effect of such an avalanche on the slip profile is shown in
\cref{fig:roughness},
illustrating the definitions of the spatial extension $A$ and the total slip $S$.
As shown in \cref{fig:PS,fig:SA,fig:TA}, we confirm powerlaw behaviours for the distribution of
avalanches $P(S)\sim S^{-\tau}$, the avalanche geometry $S\sim A^{d_f}$, and duration $T\sim A^z$;
with $\tau\approx 1.77$, $z\approx 0.64$ \footnote{
    An exponent $z<1$ is asymptotically impossible,
    since it would lead to a diverging propagation speed $v\sim A^{1-z}$ for large events.
    Once the speed of sound is reached, one presumably finds $z=1$.
    At that point, our hypothesis on the
    existence of a characteristic slip rate of avalanches must break down, instead we expect
    this rate to decrease with $A$.
    However, this limit is presumably very hard to reach empirically.
    In our system, we estimate that the speed of sound would be reached for $A=40000$ blocks
    or $2000 L_c$, far beyond what we can achieve numerically.
    This crossover is also presumably not observable in earthquakes,
    since $L_c$ is believed to be kilometric in faults.
}
and $d_f\approx 1.60$.

\paragraph{Nucleation size}

To estimate the nucleation size $A_c$ beyond which an avalanche becomes a rupture front spanning
the system, we measure the cumulative distribution of avalanche elongation $\Phi(A)$
at various stress levels, as shown in \cref{fig:PhiA}.
Next, we fit $\Phi(A) \sim A^{1 - \expkappa} \exp(- A / A_c)$ \footnote{
    $\tau_A = 1 + (\tau - 1) (1 + \zeta)$ as follows from \cref{eq:PS,eq:S,eq:df}.
}
so as to extract $A_c$.
In \cref{fig:PhiA_collapse}, we confirm that rescaling $\Phi(A)$ by such an
obtained $A_c$ indeed collapses the different curves.
There is a considerable empirical
indeterminacy on the exponent $\nu$ entering $A_c \sim (\sigma - \sigma_c)^{-\nu}$, because the
value of $\sigma_c$ is not known a priori.
To proceed, we impose the predicted value $\nu=1 / (1 - \zeta)$ with
$\zeta=0.60$ and choose $\sigma_c$ such as to obtain the best powerlaw behaviour,
as displayed in \cref{fig:PhiA_collapse}.
We obtain a good fit,
showing that $\nu$ is consistent with our prediction.
Most importantly, we get
$\sigma_c\approx 0.17\pm 0.01$ which is consistent with the value $\sigma_{\min}\approx 0.17$
extracted from the effective flow curve.
Since rate-and-state predicts a threshold for unstable
slip events at $\sigma^*\approx \sigma_{\min}$, our observations support that $\sigma^*$ indeed
controls avalanches in disordered frictional interfaces.

\paragraph{Summary of Results}

\begin{table}[htp]
    \centering
    \caption{
        Overview of results: scaling predictions and measured exponents in our $d = 1$ system.
        We report the fitted exponents in \cref{fig:zeta_all,fig:avalanche}.
        The uncertainty sums the statistical error plus an estimate of the systematic error stemming
        from the finite system size (which is estimated by considering the change in exponent in a
        twice smaller system).
    }
    \label{tab:results}
    \begin{tabular}{lll}
        \hline
        Scaling                               & Prediction                   & Measurement
        \\
        \hline
        $P(S) \sim S^{-\tau}$                 & -                            & $1.77 \pm 0.25$
        \\
        $\delta u(r) \sim r^\zeta$            & -                            & $0.60 \pm 0.08$
        \\
        $S \sim A^{\expdf}$                   & $1 + \zeta \simeq 1.60$      & $1.60 \pm 0.09$
        \\
        $T \sim A^z$                          & $\zeta \simeq 0.60$          & $0.64 \pm 0.06$
        \\
        $A_c \sim (\delta \sigma)^{- \expnu}$ & $1 / (1 - \zeta) \simeq 2.5$ & $2.25 \pm 0.77$
        \\
        \hline
    \end{tabular}
\end{table}

The set of exponents measured are reported in \cref{tab:results}.
$d_f$ and $z$ are in excellent
agreement with our predictions ($\expnu$ is nicely consistent with those, but cannot be extracted
precisely).

These exponents differ markedly from those obtained in the absence of inertia with the same
dimension $d = 1$ and long-range interactions, for which it is found numerically that
$\zeta \in [0.34, 0.39]$
\cite{Ramanathan1998,Rosso2002,Middleton1992,Narayan1993,Tanguy1998,Duemmer2007}.
These values are
in reasonable agreement with renormalization group (RG) predictions
\cite{Ertas1994,Narayan1993,Chauve2001}.
However, RG has not been successfully developed when
velocity weakening is important.
In that case, our results indicate the existence of a new
universality class.

At the experimental level,
crack propagation \cite{Schmittbuhl1997,Tallakstad2011,Vincent-Dospital2021} and
contact line experiments \cite{Prevost2002,Rubio1989} often report $\zeta \in [0.5,0.7]$.
These exponents are closer to our predictions, yet it is unclear if inertia is
responsible for this discrepancy with over-damped numerical observations
\cite{Ramanathan1998,Rosso2002,Middleton1992,Narayan1993,Tanguy1998,Duemmer2007},
or if other effects are at play \cite{Vincent-Dospital2021}.

\section{Conclusion \& perspective}

\paragraph{Summary}

We have introduced a theoretical framework for the nucleation and statistics of slip
at a disordered frictional interface.
It builds on rate-and-state results \cite{Zheng1998,Brener2018}
showing that in the presence of strong velocity-weakening effects,
a homogeneous system presents a threshold stress $\sigma^*$ beyond which
a rupture can invade the system.
We have argued that in the presence of disorder,
such a threshold must lead to powerlaw avalanches.
Rupture occurs when one avalanche becomes larger than some size $A_c$ beyond which the
disorder becomes negligible and cannot stop a rupture.
$A_c$ diverges as $\sigma \rightarrow \sigma^*_+$ with some new exponent.
This framework leads to quantitative predictions, partly based on extending arguments from the
depinning literature \cite{Fisher1998}, where the threshold stress can be defined statically, to
situations where the threshold is dynamically defined.
Most importantly, our theoretical approach should stand as long as the frictional interface is
well described by rate-and-state,
irrespectively of the underlying mechanism causing velocity-weakening.
Note that we have numerically checked this prediction only in a specific model,
leaving a check to broader classes of models for future work.

Next, we used a minimal model of frictional interface as a Rosetta Stone, in which
(i) rate-and-state equations can be calibrated and their predictions tested and
(ii) disorder is easily controllable and slip statistics readily measurable.
It allowed for a stringent test of our scaling predictions, and put forwards numerical values for
exponents that future theories should seek to explain.

\paragraph{Geophysical data}

We have argued that for a disordered but overall flat velocity-weakening frictional interface,
the distribution of slip events should be bimodal.
Powerlaw distributed avalanches are present, which display scaling properties.
We find for example a scaling relationship between the avalanche size or ``seismic moment''
\footnote{
    The seismic moment is defined as the average slip times the slipping area,
    times the shear modulus.
    Our definition of $S$ is a proxy for the former product,
    taking that the slip of a block is proportional to the number of times the block yields,
    and assuming that $A$ is a proxy for the linear extension of the avalanche which
    we justify by the observation that avalanches a compact
    -- each block yields many times during an avalanche, see \cref{fig:eventmap}.
}
$S$ and its associated stress drop $\delta \sigmaf \sim S^{-1 / (d_f \nu)} \sim S^{-0.25}$.
In the geophysics literature, these empirical facts are debated.
Some studies report that the stress drop mildly decreases \cite{Shearer2006} or increases
\cite{Trugman2017} with the earthquake size.
Furthermore, as noted in the introduction,
for a single fault bimodal \cite{Wesnousky1994} or mono-modal \cite{Page2011} slip distributions
have been argued for.
In our opinion, the interpretation of these results is complicated by the geometry of faults, which
display broadly distributed segments where slip can occur \cite{Manighetti2009}.
We view our theoretical results as a first step focusing on a simple interface geometry.
Arguably, more complex geometrical factor must be included to rationalise geophysical observations.
In this respect, it would be very interesting to consider a frictional interface made of powerlaw
distributed segments.
It can be readily implemented in our model, where the geometry of plastic regions where slip occurs
can be chosen at will.

\paragraph{Future works}

More generally,
our methodology corresponds to a minimal model with a desired rate-and-state behaviour.
In the future, controlled disorder can be used to incorporate other phenomena of interest,
and study how they shape slip statistics.
A particularly relevant case is thermal creep, which is expected to lead to an $N$-shaped effective
flow curve, which can be readily obtained by simulating the dynamics in our model at finite
temperature.
Likewise, the materials surrounding the interface can be viscoelastic instead of elastic, which is
a sufficient condition (but not always necessary \cite{Houdoux2021}) to obtain aftershocks
\cite{Jagla2014}.

\section*{Acknowledgement}

We thank A.~Rosso, J.-F.~Molinari, T.D.~Roch, M.A.D.~Lebihain, M.~Popovi{\'c}, E.~Agoritsas, W.~Ji,
M.~Violay
and M.~M{\"u}ller for discussions, and J.~Volmer and F.~Barras for feedback on the manuscript.
T.G.~acknowledges support from the Swiss National Science Foundation (SNSF) by the SNSF Ambizione
Grant PZ00P2{\_}185843.
M.W acknowledges support from the Simons Foundation Grant (No.~454953 Matthieu Wyart) and from the
SNSF under Grant No.~200021-165509.

\bibliographystyle{unsrtnat}
\bibliography{library}

\appendix

\section{Details of the model}

\paragraph{Equation of motion}

We consider standard continuum elasto-dynamics, so that the equation of motion reads
\begin{equation}
    \tilde{\rho} \partial_{\tilde{t}}^2 \tilde{w}_i (r) =
    \mathrm{div} \big( \tilde{\sigma}_{ij} (r) \big) -
    \tilde{\alpha} \partial_{\tilde{t}} \tilde{w}_i (r)
\end{equation}
where $\tilde{w}_i (r)$ is the displacement field
(a function of position $r$, whose vectorial nature is omitted for notational simplicity),
and $\partial_{\tilde{t}} \tilde{w}_i$ and $\partial_{\tilde{t}}^2 \tilde{w}_i$ its first and
second time derivative.
$\tilde{\rho}$ is the mass density and $\tilde{\alpha}$ is the (small) damping coefficient,
both are taken homogeneous.
$\mathrm{div} \big( \tilde{\sigma}_{ij} (r) \big)$ is the divergence of the stress tensor
$\tilde{\sigma}_{ij}$.
The latter follows from linear elasticity, that we model
\begin{equation}
    \begin{aligned}
        \tilde{\sigma}_{ij} (r) & =
        (\tilde{\kappa} / (d + 1)) \mathrm{tr} \big( \tilde{\varepsilon}_{ij} (r) \big) \delta_{ij}      \\
                                & + 2 \tilde{\mu} (\tilde{u} (r) - \tilde{u}_\text{min} (r)) N_{ij} (r).
    \end{aligned}
\end{equation}
Here, $\tilde{\varepsilon}_{ij}(r) = (\partial_i \tilde{w}_j(r) + \partial_j \tilde{w}_i(r)) / 2$
is the strain tensor.
$d + 1$ \footnote{
    We use $d$ as the dimension of the interface.
} is the dimension of the bodies (here $d + 1 = 2$), $\tilde{\kappa}$ is the bulk modulus,
$\tilde{\mu}$ is the shear modulus;
$\delta_{ij}$ is the unit tensor and
$\mathrm{tr}(a_{ij}) = a_{ij} \delta_{ij}$ is the trace of $a_{ij}$.
$N_{ij}(r)$ defines the direction of shear as
\begin{equation}
    N_{ij}(r) = \mathrm{dev}(\tilde{\varepsilon}_{ij} (r)) / \tilde{u}(r)
\end{equation}
where $\mathrm{dev}(\tilde{\varepsilon}_{ij}(r))$ is the deviatoric (trace-free) part
of the strain tensor and
\begin{equation}
    \label{eq:equi_strain}
    \begin{aligned}
        \tilde{u}(r) & \equiv || \tilde{\varepsilon}_{ij}(r) ||_d \\
                     & \equiv
        (\mathrm{dev}(\tilde{\varepsilon}_{ij}(r))\mathrm{dev}(\tilde{\varepsilon}_{ij}(r))/2)^{1 / 2}
    \end{aligned}
\end{equation}
corresponds to the magnitude of the shear strain, that we refer to as ``slip''.
Likewise, we define the magnitude of shear stress
\begin{equation}
    \label{eq:equi_stress}
    || \tilde{\sigma}_{ij}(r) ||_d \equiv
    (2 \, \mathrm{dev}(\tilde{\sigma}_{ij}(r))\mathrm{dev}(\tilde{\sigma}_{ij}(r))^{1 / 2}.
\end{equation}

The potential energy landscape in \cref{fig:model} is defined along $\tilde{u}(r)$, with
$\tilde{u}_\text{min}(r)$ the currently closest local minimum along the coordinate $\tilde{u}(r)$.
It is always equal to zero in the bulk (in blue in \cref{fig:model}),
but typically finite along the `weak' layer (in red in \cref{fig:model}).
Along that layer, the cusps are separated by a distance chosen randomly from a Weibull
distribution, that has a typical value $2 \tilde{u}_0$.

\paragraph{Units}

A typical magnitude of shear strain is the typical yield strain $\tilde{u}_0$ of a block,
that we use to define units,
such that $\varepsilon_{ij}(r) \equiv \tilde{\varepsilon}_{ij}(r) / \tilde{u}_0$
and $\sigma_{ij}(r) \equiv \tilde{\sigma}_{ij}(r) / \tilde{\sigma}_0$
with $\tilde{\sigma}_0 \equiv 4 \tilde{\mu} \tilde{u}_0$.
Thereby, we denote dimensionless quantities $\bullet$ and their dimension-full equivalent
$\tilde{\bullet}$.

We define the plastic slip $u_p(r) \equiv \tilde{u}_{\min}(r) / \tilde{u}_0$
as the location of the current local minimum in dimensionless slip space, see \cref{fig:model}.
These definitions are such that, on average, the number of times a block yields
$s(r) = \Delta u_p(r) / 2 \approx \Delta u(r) / 2$.
The slip rate $\dot{u} \equiv \Delta u / \Delta t$,
with time $t \equiv \tilde{t} / \tilde{t}_0$, where
$\tilde{t}_0 \equiv \ellblocktilde / \tilde{c}_s$ with $\tilde{c}_s$ the shear wave speed.
We note that length is expressed in units of $\ellblocktilde$ such that
$L \equiv \tilde{L} / \ellblocktilde$ and $\ellblock \equiv \ellblocktilde / \ellblocktilde = 1$.
In our dimensionless units,
slip at the interface, that we define as the strain in the blocks,
thus coincides with the displacement discontinuity across the interface.
Furthermore, time $t$ indicates the number of blocks a shear wave traversed.
A slip rate $\dot{u}(r) = 0.5$ thus indicates that a typical block yields once during time
it takes a shear wave to travel the distance of one block.

We extract the total slip $S \equiv \int_L s(r) dr$
($\int_L \ldots dr$ denotes the integral along the weak layer)
as the total number of times
blocks yield, $A$ the number of blocks that yield at least once
(thus $A \equiv \int_L (s(r) + |s(r)|) / (2 |s(r)|) dr$),
and $T$ the (dimensionless) duration between the first and the last time that a block yield
during an event.

\paragraph{Numerical model}

The numerical treatment of this equation of motion corresponds to a discretisation in space
using finite elements (where at the weak layer the elements coincide with the blocks
of linear size $\tilde{\ell}_0$, see \cref{fig:model}),
and in time using the velocity Verlet algorithm.
The numerical values of all parameters, and more details,
can be found in \cite{deGeus2019,deGeus2019_data}.
Different from those references, here we consider a bigger system of $L = 4 \times 3^6$ blocks
(except for the results in \cref{fig:eventmap} which are made on the system of \cite{deGeus2019}),
and a ten times smaller typical strain $\tilde{u}_0$
to acquire more events per realisation while respecting the small strain assumption.
Note that this does not lead to any change in terms of the dimensionless quantities
reported here and in \cite{deGeus2019}.
In addition, we perform flow experiments by imposing a fixed shear rate
to the top boundary.
In practice, the shear is supplied to the system in a distributed manner,
such that in each time step all nodal displacements are updated
according to an affine simple shear, though only the top boundary is fixed.
We measure both $\sigmaf$ and $\dot{u}$ as
averaged in space along the interface and on a finite window of time deep in the steady state,
as well as on different realisations.

\paragraph{Quantities}

The remote stress is the volume averaged shear stress
\begin{equation}
    \sigma \equiv
    \left|\left|
    \iint_L \bm{\sigma}(\vec{r}) \, d\vec{r} \,
    \right|\right|_d =
    \sum_{\beta = 1}^n (\tilde{f}_x)_\beta / (n \tilde{\ell}_0 \tilde{\sigma}_0)
\end{equation}
with $\bm{\sigma}(\vec{r})$ the adimensional stress tensor at a position
$\vec{r}$ in $(d + 1)$-dimensional space,
and $\iint_L \ldots dr$ the integral over the entire domain in $(d + 1)$-dimensional space.
$(\tilde{f}_x)_\beta$ are the reaction forces in horizontal direction
of the $n + 1$ nodes along the top boundary
(one node is `virtual' because of
the periodic boundary conditions in horizontal direction) whose position is prescribed.
The remote strain is the volume averaged strain
\begin{equation}
    \varepsilon \equiv
    \left|\left|
    \iint_L \bm{\varepsilon}(\vec{r}) \, d\vec{r} \,
    \right|\right|_d =
    (\tilde{w}_x)_\beta / (\tilde{H} \tilde{u}_0)
\end{equation}
with $(\tilde{w}_x)_\beta$ the displacement in horizontal direction of one of the nodes along
the top boundary (the displacement of all of these nodes is definition equal),
and $\tilde{H} \approx L \tilde{\ell}_0$ the actual height of the sample.

The stress along the interface
\begin{equation}
    \sigmaf \equiv
    \left|\left|
    \int_L \bm{\sigma}(r) \, dr \,
    \right|\right|_d = \left|\left|
    \sum_{i = 1}^L \bm{\sigma}_i \,
    \right|\right|_d
\end{equation}
with $i$ referring the block index along the weak layer
(numbered from left to right).
We note that
\begin{equation}
    \sigmaf (A) = \left|\left|
    \sum_{i = j}^{i + A} \bm{\sigma}_i \,
    \right|\right|_d.
\end{equation}
(where periodicity implies $\bm{\sigma}_i = \bm{\sigma}_{i + L}$).

The slip along the interface
\begin{equation}
    u \equiv
    \left|\left|
    \int_L \bm{\varepsilon}(r) \, dr \,
    \right|\right|_d = \left|\left|
    \sum_{i = 1}^L \bm{\varepsilon}_i \,
    \right|\right|_d.
\end{equation}
Finally, the slip rate
\begin{equation}
    \dot{u} \equiv
    \left|\left|
    \int_L \, \partial_t \bm{\varepsilon}(r) \, dr \,
    \right|\right|_d = \left|\left|
    \sum_{i = 1}^L \partial_t \bm{\varepsilon}_i \,
    \right|\right|_d.
\end{equation}

\paragraph{Radiation damping}

A nucleating event,
whereby part of the interface and bulk are still static as the rupture invades the interface,
is stabilised by the bulk surrounding it:
to increase the slip rate $\dot{u}$ the bulk around the rupture has to be accelerated.
Due to the cost of accelerating an expanding volume,
the interfacial stress $\sigmaf$ inside the event differs from the remote stress $\sigma$.
Because the bulk is accelerated by elastic waves that radiate away from the interface
this effect is commonly referred to as ``radiation damping''.
We emphasise that this is an effect of standard elasto-dynamics:
it is not added by hand to our model.

The effect of radiation damping corresponds to a `cost' of stress \cite{Zheng1998}
$\Delta \tilde{\sigma}_{xy} = \tilde{v} \tilde{\mu} / (2 \tilde{c}_s)$,
with $\tilde{v} \equiv 2 \Delta (\delta \tilde{r}_x) / \Delta \tilde{t}$ \cite{Barras2020}
the rate of change of the displacement discontinuity
$\delta \tilde{r}_x
    \equiv 2 \tilde{\varepsilon}_{xy} \ellblocktilde = 2 \tilde{u} \ellblocktilde$
such that
$\Delta \tilde{\sigma}
    = (4 \Delta \tilde{u} \ellblocktilde / \Delta \tilde{t}) (\tilde{\mu} / \tilde{c}_s)
    = 4 \tilde{\mu} \Delta \tilde{u} / (\Delta \tilde{t} / \tilde{t}_0)$
or
$\Delta \tilde{\sigma} / (4 \tilde{\mu} \tilde{u}_0)
    = (\Delta \tilde{u} / \tilde{u}_0) / (\Delta \tilde{t} / \tilde{t}_0)$
and thus
$\Delta \sigma = \dot{u}$.

\paragraph{Computation of $L_c$}

For the rate-and-state law
$\tilde{L}_c
    = - \pi \tilde{\mu} \tilde{D}_c /
    (\dot{\tilde{u}} \partial \tilde{\sigma} / \partial \dot{\tilde{u}})$
\cite{Brener2018} with
$\partial \tilde{\sigma} / \partial \dot{\tilde{u}}
    = (\tilde{a} - \tilde{b}) / \dot{\tilde{u}}$
the derivative of the steady state in \cref{eq:rate-and-state}, such that
$\tilde{L}_c = - \pi \tilde{\mu} \tilde{D}_c / (\tilde{a} - \tilde{b})$.
In our model, a block, on average, loses memory over a sliding distance
$\tilde{D}_c = \tilde{u}_0 \ellblocktilde$.
Using, furthermore, our units of stress such that $a = \tilde{a} / \tilde{\sigma}_0$ and
$b = \tilde{b} / \tilde{\sigma}_0$, we find
$\tilde{L}_c
    = - \pi \tilde{\mu} \tilde{u}_0 \ellblocktilde / (\tilde{\sigma}_0 (a - b))
    = - \pi \ellblocktilde / (4 (a - b))$,
and thus
$L_c = - \pi / (4 (a - b)) \approx 26$.

\paragraph{Powerlaw fits}

The powerlaw fit of $y = c x^b$ is performed using a least square fit of the linear relation
$z \equiv \ln y = \ln c + b \ln x$.
In the case of an uncertainty $\delta y$ (typically a standard deviation)
we assume that $\delta y \ll y$ such that we use $\delta z = \delta y / y$.
The error of the fitted exponent, $\delta b_f$, is then the square root of the relevant component
at the diagonal of the 2x2 covariance matrix.
Where possible, we also compute the fluctuations of the exponent, $\delta b_\ell$,
by reducing the fitting range by a factor of two and four.
We report $\delta b_\ell$ in \cref{fig:zeta}, and
$\delta b_f$ in \cref{fig:PS,fig:SA,fig:TA}
(in \cref{fig:PS,fig:SA} the $\delta b_\ell$ was simply found lower or equal to $\delta b_\ell$;
in \cref{fig:TA} the range is not sufficient to be reduced).

In \cref{fig:PhiA_collapse} we account for the error in $\sigma$
by taking the error as dimensionless, allowing us to compose it
in equal amounts of the errors in $\sigma$
(the standard deviation of $\sigma$ in each bin)
and in $A_c$ (the fitting error from \cref{fig:PhiA}).
We use this protocol also to fit $\nu$ given $\sigma_{\min} \approx 0.17$
as reported in \cref{tab:results}.

In \cref{tab:results} we estimate the error on $\nu$ as the difference between our prediction
and a fit of the exponent of the data in \cref{fig:PhiA_collapse} using $\sigma_c$ defined as
the bottom of the effective flow curve in \cref{fig:scenario}.

\end{document}